\documentclass[sn-nature, iicol]{sn-jnl}

\usepackage{graphicx}%
\usepackage{multirow}%
\usepackage{amsmath,amssymb,amsfonts}%
\usepackage{amsthm}%
\usepackage{mathrsfs}%
\usepackage[title]{appendix}%
\usepackage{xcolor}%
\usepackage{textcomp}%
\usepackage{manyfoot}%
\usepackage{booktabs}%
\usepackage{algorithm}%
\usepackage{algorithmicx}%
\usepackage{algpseudocode}%
\usepackage{listings}%
\usepackage{physics}%
\usepackage{mathtools}%
\usepackage{subcaption}%
\usepackage{siunitx}
\usepackage{bm}

\unnumbered

\begin{document}

\title[Quantum Quench Dynamics of Geometrically Frustrated Ising Models]{Quantum Quench Dynamics of Geometrically Frustrated Ising Models}


\author[1]{\fnm{Ammar} \sur{Ali}}\equalcont{These authors contributed equally to this work.}

\author[2]{\fnm{Hanjing} \sur{Xu}}
\equalcont{These authors contributed equally to this work.}

\author[3]{\fnm{William} \sur{Bernoudy}}
\author[4, 5]{\fnm{Alberto} \sur{Nocera}}
\author*[3]{\fnm{Andrew D.} \sur{King}}\email{aking@dwavesys.com}
\author*[1]{\fnm{Arnab} \sur{Banerjee}}\email{arnabb@purdue.edu}

\affil[1]{\orgdiv{Department of Physics and Astronomy}, \orgname{Purdue University}, \orgaddress{\city{West Lafayette}, \state{IN}, \postcode{47906}, \country{USA}}}
\affil[2]{\orgdiv{Department of Computer Science}, \orgname{Purdue University}, \orgaddress{\city{West Lafayette}, \state{IN}, \postcode{47906}, \country{USA}}}
\affil[3]{\orgname{D-Wave}, \orgaddress{\city{Burnaby}, \state{British Columbia}, \country{Canada}}}
\affil[4]{\orgdiv{Stewart Blusson Quantum Matter Institute}, \orgname{University of British Columbia}, \orgaddress{\city{Vancouver}, \state{British Columbia}, \postcode{V6T1Z4}, \country{Canada}}}
\affil[5]{\orgdiv{Department of Physics Astronomy}, \orgname{University of British Columbia}, \orgaddress{\city{Vancouver}, \state{British Columbia}, \postcode{V6T1Z1}, \country{Canada}}}


\abstract{Geometric frustration in two-dimensional Ising models allows for a wealth of exotic universal
behavior, both Ising and non-Ising, in the presence of quantum fluctuations. In particular, the triangular antiferromagnet and Villain model in a transverse field can be understood through
distinct XY pseudospins, but have qualitatively similar phase diagrams including a quantum
phase transition in the (2+1)-dimensional XY universality class. While the quantum dynamics of
modestly-sized systems can be simulated classically using tensor-based methods, these methods
become infeasible for larger lattices. Here we perform both classical and quantum simulations
of these dynamics, where our quantum simulator is a superconducting quantum annealer. Our
observations on the triangular lattice suggest that the dominant quench dynamics are not described by the quantum
Kibble-Zurek scaling of the quantum phase transition, but rather a faster coarsening dynamics in
an effective two-dimensional XY model in the ordered phase. Similarly, on the Villain model, the scaling exponent does not match the Kibble-Zurek expectation. These results demonstrate the ability of quantum annealers to simulate coherent quantum dynamics and scale beyond the reach of classical approaches.}





\maketitle

\section{Main}\label{sec1}

Frustrated systems are systems in which it is impossible to simultaneously minimize all Hamiltonian terms, either due to the non-commutativity of the terms or the lattice geometry (geometric frustration). Such models are of great interest in both classical and quantum spin systems due to their ability to maintain topological order, by giving rise to topologically protected quasi-particle excitations, and host topological defects, such as 1D kinks, 2D vortices, domain walls, etc. \cite{SCHMIDT20171, Shaginyan_Stephanovich_Msezane_Japaridze_Clark_Amusia_Kirichenko_2019, Moessner2001, anderson1973, Savary2017, DefectReview}.\\

\begin{figure}[t]
    \centering
    \hspace{-1cm}
    \includegraphics[scale = 0.16]{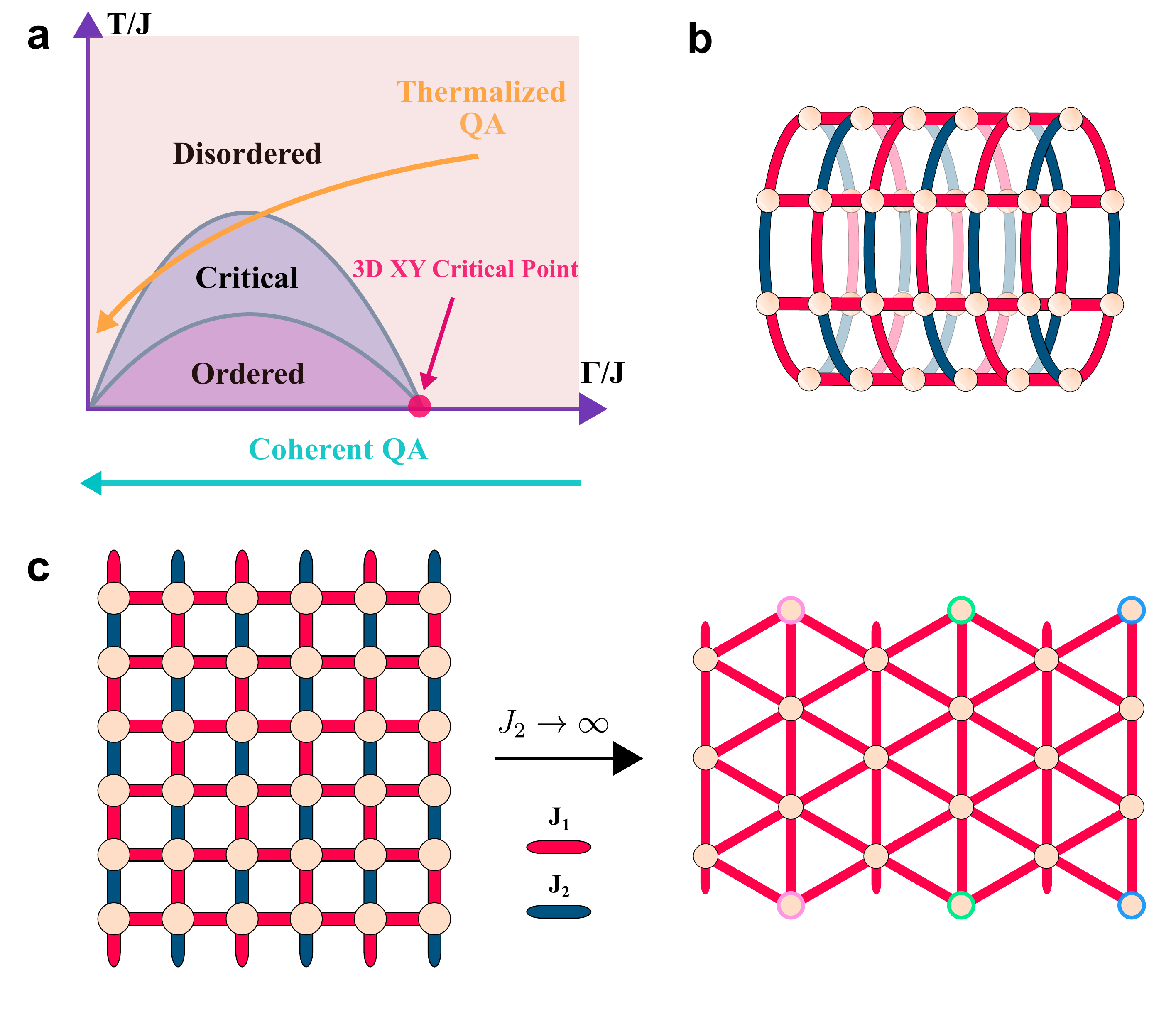}
    \caption{\textbf{Phase diagram and lattice definition.} \textbf{a,} Qualitative phase diagram of the triangular and Villain TFIMs (adapted from \cite{Moessner2001}). The orange arrow shows the KT phase transition explored in \cite{King2018}, while the cyan arrow depicts the path considered in this work, passing through the 3D XY quantum critical point. \textbf{b,} Embedded lattice on the quantum annealer (QA) corresponds to a square lattice with periodic boundary conditions along one direction giving a cylinder. \textbf{c,} Shown is the choice of couplings and the contraction to an AFM triangular lattice when taking the magnitude of the FM couplers ($J_2$) to be much larger than the AFM ones ($J_1$). Nodes sharing the same border color are identified.}
    \label{fig1}
\end{figure}

\noindent
Focusing on geometric frustration and its defects, two closely related models of particular relevance to quantum systems are classical stacked frustrated magnets: the 3D antiferromagnetic (AFM) triangular Ising model \cite{Blankschtein1984a}, and the 3D frustrated Ising simple cubic lattice \cite{Blankschtein1984}, both of which are known to undergo thermal phase transition that belongs to the 3D XY universality class \cite{Blankschtein1984a, Blankschtein1984}. By virtue of the classical-quantum duality \cite{sachdev1999}, the 2D quantum versions of these systems are expected to have quantum phase transitions belonging to the same universality class \cite{Jalabert1991}. Indeed, the 2D Transverse Field Ising Model (TFIM) on an AFM triangular lattice turned out to have a quantum critical point in the 3D XY universality class (Fig. \ref{fig1}a) \cite{Moessner2000, Moessner2001, Isakov2003}. Similarly, the TFIM on a fully frustrated square lattice, known as the Villain model \cite{villain1977}, was shown to share the same critical behavior \cite{Moessner2001}. \\

\noindent
Simulation of a frustrated system on a quantum annealer (QA) was realized in \cite{King2018} where the thermal Kosterlitz-Thouless (KT) phase transition \cite{kosterlitz1973} was observed. Correlations between the topological defects---vortices and antivortices---were shown to have an exponential decay above the critical temperature and a power-law decay below it, showcasing the binding of defects into vortex-antivortex pairs. In \cite{King2021}, it was shown that simulating finite-temperature frustrated quantum systems on QAs greatly accelerates the computational performance where a speedup up to $10^6$ was reported over Path-Integral Monte Carlo simulation (PIMC), a leading classical method \cite{King2021}. In the same vein, it was lately shown in \cite{bauza2024scaling} that approximate optimization on QAs has a scaling advantage over the classical heuristic algorithm parallel tempering with isoenergetic cluster moves (PT-ICM). More recently, coherent quantum annealing has been demonstrated in superconducting processors, allowing quantum simulation of programmable geometries on thousands of qubits with negligible coupling to the thermal environment.\\
\begin{figure*}[!htbp]
    \centering
    \includegraphics[scale = 0.2]{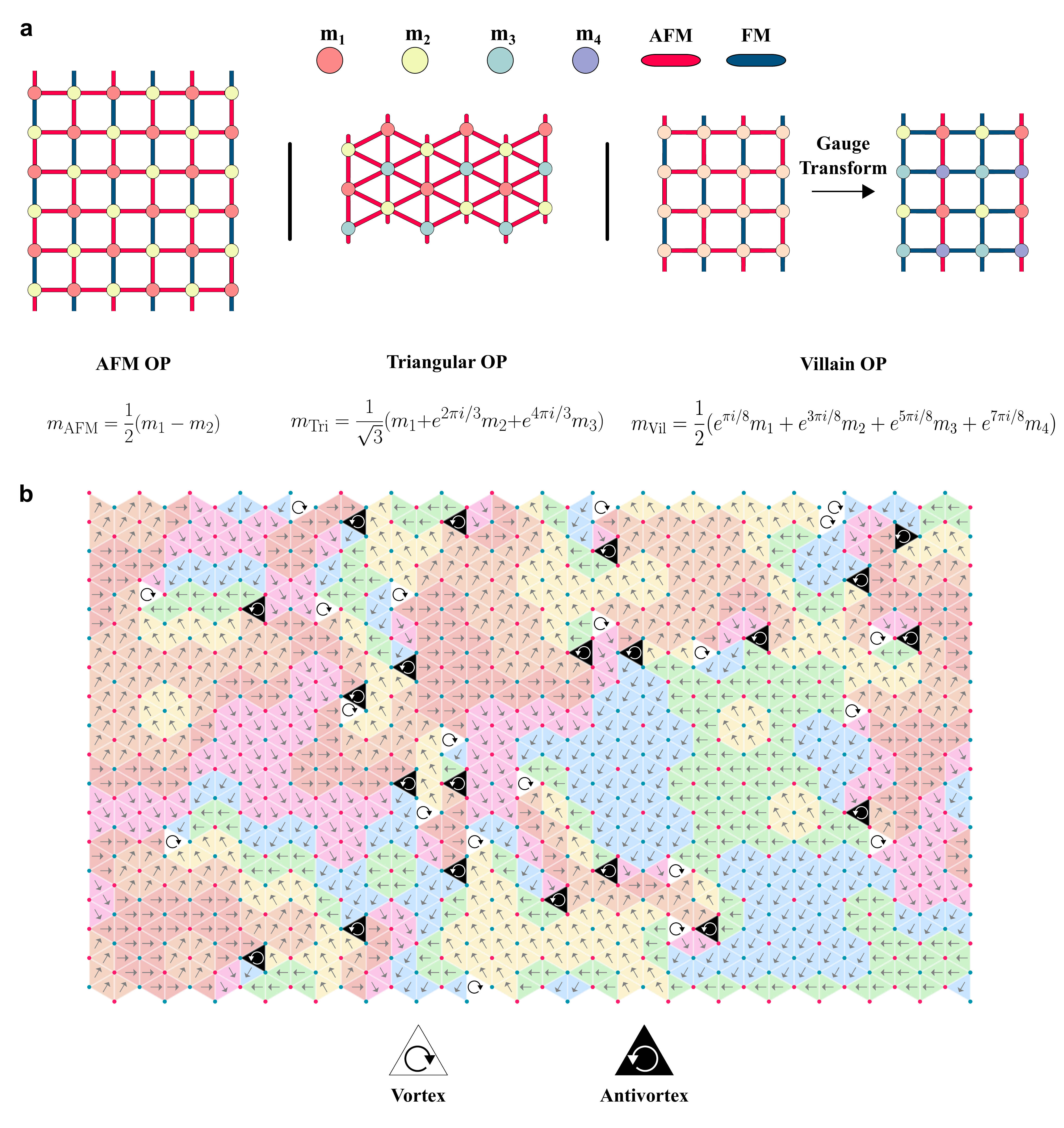}
    \caption{\textbf{Order parameters (OPs) definitions and topological defects.} \textbf{a,} Pictorial representation of the three different OPs, given in equations \ref{eq4}, \ref{eq5}, \ref{eq6}. \textbf{b,} A QA output instance for $L = 36$, and $\text{t}_a = 6.3$ns. The pseudospin of each plaquette is indicated giving rise to the topological defects represented by the vortices (white) and antivortices (black).}
    \label{fig2}
\end{figure*}

\noindent
Coherent quantum annealing was first realized in \cite{King2022} where they studied the quenching of a 1D TFIM chain. It was shown that the defect density $\rho$ resulting from the equilibration of the chain into alternating $\mathbb{Z}_2$ ordered domains follows the Kibble-Zurek mechanism (KZM) scaling \cite{kibble1976, zurek1985}

\begin{equation}\label{eq1}
    \rho \propto \text{t}_a^{\frac{-d\nu}{1+z\nu}},
\end{equation}
where $\text{t}_a$ is the annealing time, $d$ is the space dimensionality, z and $\nu$ are the dynamical and correlation length critical exponents of the 2D Ising universality class, respectively. KZM was also observed on a 1D Rydberg atom array in \cite{Keesling2019}, where also $\mathbb{Z}_3$ and $\mathbb{Z}_4$ broken-order phases were explored, and on a 2D Rydberg atom square array in \cite{Ebadi2021}. Further coherent quantum dynamics were observed in a 3D spin glass system where a scaling advantage of quantum annealing over both simulated annealing and simulated quantum annealing was reported \cite{King2023}.\\

\noindent
In this work, we explore the quenching of 2D frustrated systems, namely the AFM TFIM triangular and Villain models. Fig. \ref{fig1}b, c show the considered lattice, a square lattice of size $L\times L$ with periodic boundary conditions along the vertical direction. The D-Wave Advantage QA realizes the TFIM Hamiltonian
\begin{equation}\label{eq2}
    \mathcal{H}(s) = -\Gamma(s) \sum_i \sigma_i^x + \mathcal{J}(s) \sum_{<ij>} J_{ij} \sigma_i^z \sigma_j^z,
\end{equation}


\noindent
where $s$ is the normalized time $\text{t}/\text{t}_a$ going from $0$ to $1$. The parameters $\Gamma(s)$ and $\mathcal{J}(s)$ evolve monotonically such that $\Gamma(0) = \mathcal{J}(1) = 1$ and $\Gamma(1) = \mathcal{J}(0) = 0$. We denote by the triangular lattice the choice of couplers $J_1 = +0.9$, $J_2 = -2$, and by the Villain model the choice $J_1 = -J_2 = 0.9$. Fig. \ref{fig1}c shows how such a choice approximates the square lattice as a triangular one by contracting every other pair of physical qubits into one logical qubit. \\
\noindent
Initially, all qubits are prepared in the superposition state
\begin{equation}\label{eq3}
   \ket{\rightarrow} = \frac{1}{\sqrt{2}} \ket{\uparrow} + \frac{1}{\sqrt{2}} \ket{\downarrow},
\end{equation}
which is the ground state of the initial Hamiltonian $\mathcal{H}(0)$. Evolving slowly enough from this ground state, the adiabatic theorem guarantees that we always stay at the instantaneous ground state. However, by quenching the system we generally end up in an excited state, whose scaling with quenching time depends on the underlying mechanism governing the dynamics, e.g. KZM or coarsening dynamics. We aim to characterize the scaling of these frustrated systems by quantifying the scaling of the order parameters (OPs) (Fig. \ref{fig2}a), defect density (Fig. \ref{fig2}b), and correlation lengths.\\

\noindent
The three OPs $m_{\text{AFM}}$, $m_{\text{Tri}}$, and $m_{\text{Vil}}$ are defined by dividing the lattice into two, three, and four sublattices, respectively, as depicted in Fig. \ref{fig2}a. They are given by (\cite{Moessner2000,schumm2023})
\begin{equation}\label{eq4}
    m_{\text{AFM}} = \frac{1}{2}(m_1 - m_2),
\end{equation}
\begin{equation}\label{eq5}
    m_{\text{Tri}} = \frac{1}{\sqrt{3}} (m_1 + e^{2\pi i/3} m_2 + e^{4 \pi i/3} m_3),
\end{equation}
\begin{equation}\label{eq6}
    m_{\text{Vil}} = \frac{1}{2}(e^{\pi i/8}m_1 + e^{3 \pi i/8}m_2 + e^{5 \pi i/8}m_3 + e^{7 \pi i/8}m_4),
\end{equation}
where $m_i$, $i \in 1, 2, 3, 4$, is the magnetization of the $i$th sublattice. We note that due to their sublattice structure, triangular (Villain) OPs are only defined for lattice dimensions multiple of three (four).\\

\begin{figure}[!htbp]
    \centering
    \hspace*{-0.3cm}
    \includegraphics[scale = 0.25]{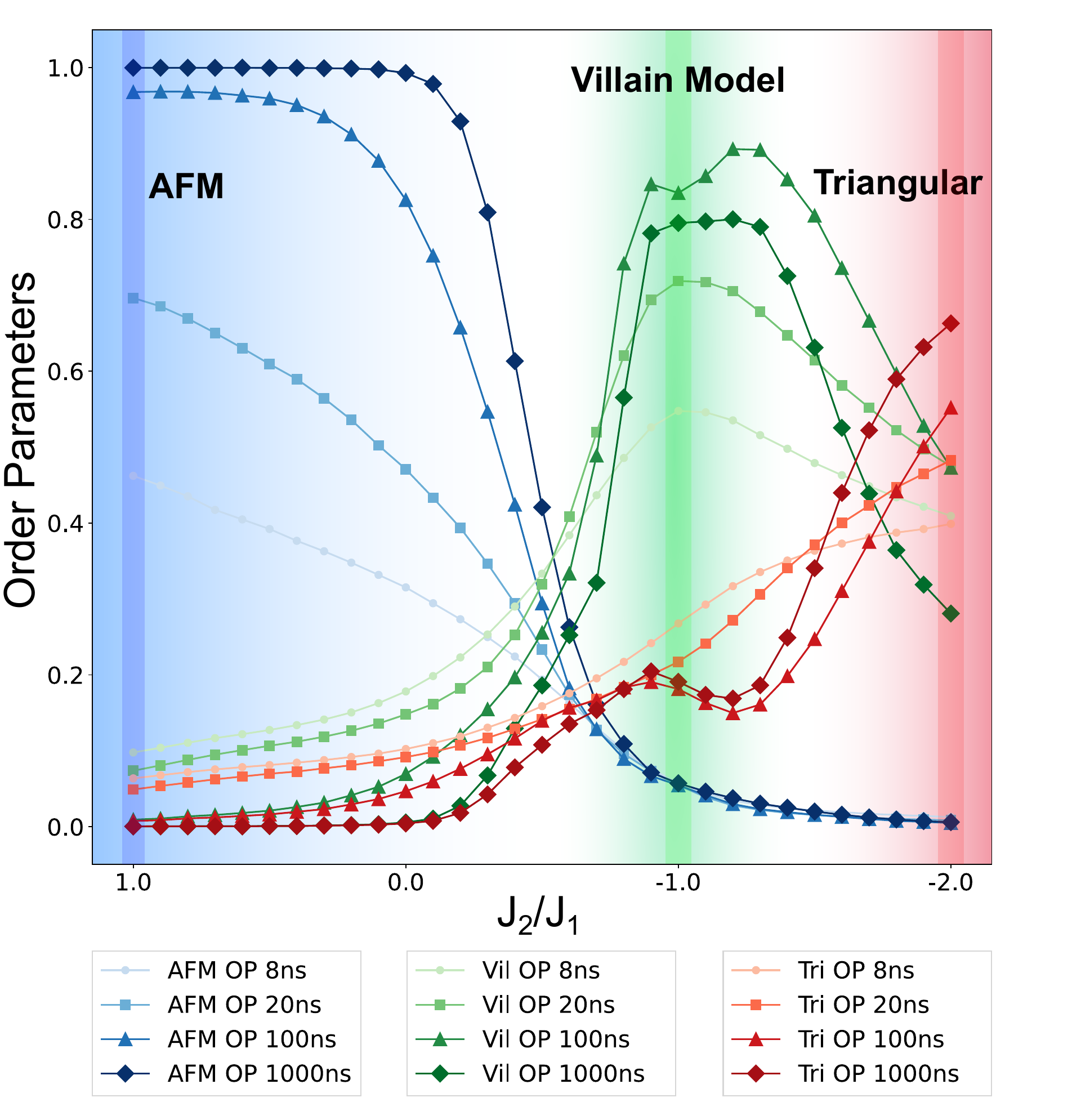}
    \caption{\textbf{OPs as a function of the ratio $J_2/J_1$.} Evolution of the different OPs evaluated on a 12$\times$12 lattice while fixing $J_1$ to be $+0.9$ and varying $J_2$ to span the ratio range from $+1$ (AFM square lattice) to $-2$ (triangular lattice approximation). We observe different regimes in which each OP is maximum. We also observe how the OPs change with changing total annealing time where we note that generally, increasing annealing time leads to increased order, i.e. approaching the classical ground state. But interestingly, this is not the case for the Villain OP indicating the influence of quantum fluctuations on its ordering.}
    \label{fig3}
\end{figure}

\begin{figure*}[!htbp]
    \centering
    \hspace*{-1cm}
    \includegraphics[scale = 0.25]{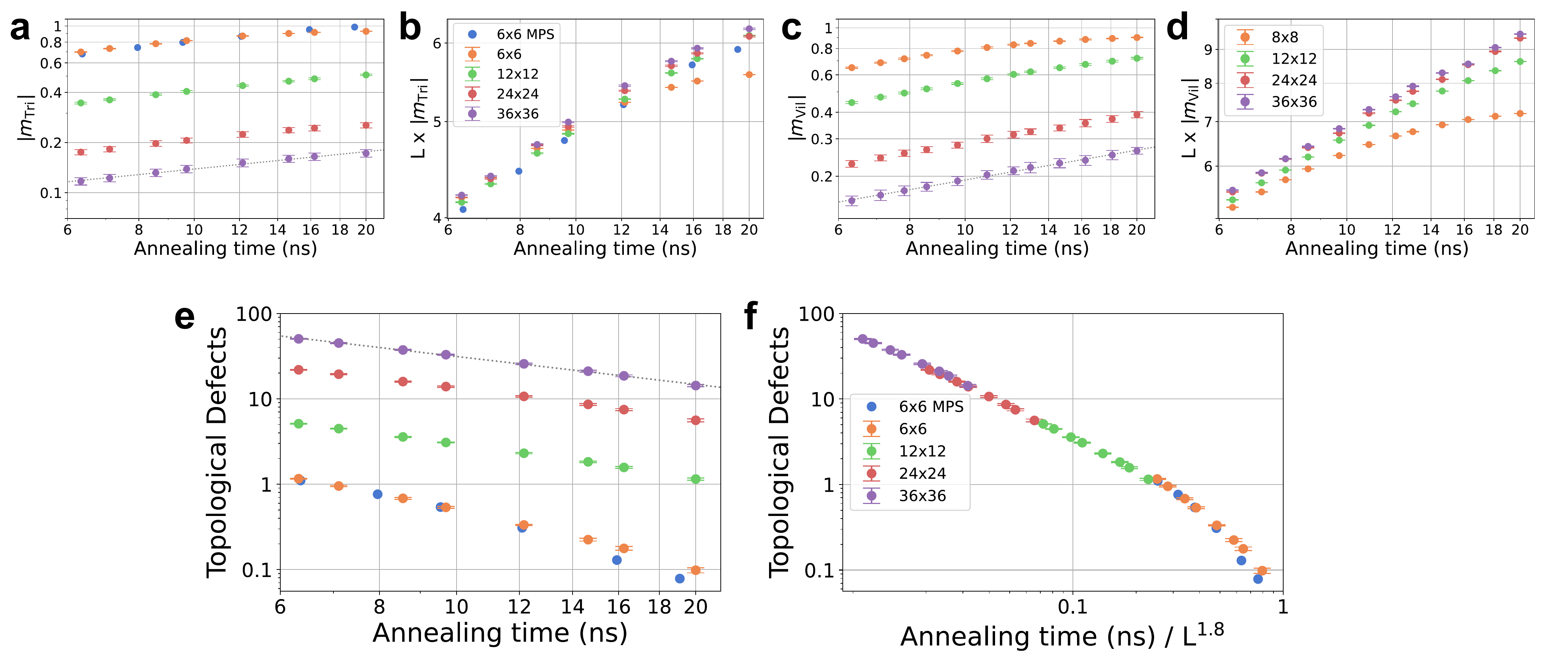}
    \caption{\textbf{Scaling and collapse of order parameters and defect count.} \textbf{a, b, } Triangular OP scaling with annealing time, and its collapse upon scaling the y-axis by the system size L. \textbf{c, d,} Same as \textbf{a, b} for the Villain OP. \textbf{e,} Scaling of the number of topological defects (vortices and anti-vortices) with annealing time. \textbf{f,} Horizontal collapse of the topological defects number, where the transition from linear to non-linear indicates a change in the underlying mechanism, as explained in the main text.}
    \label{fig4}
\end{figure*}
\noindent
Fig. \ref{fig3} probes the evolution of the three OPs on a $12\times 12$ lattice in which we fix the $J_1$ couplers to $+0.9$ and allow $J_2$ to go from $J_2 = J_1$, i.e. square AFM lattice, to $J_2 = -2J_1$, which is the triangular approximation as shown in Fig. \ref{fig1}c. We perform this for increasing values of annealing times $\SI{8}{ns}$, $\SI{20}{ns}$, $\SI{100}{ns}$, and $\SI{1}{\mu s}$. We identify different regions in the parameter space in which each OP is maximum, highlighting the phase of the system. Due to the hardware constraints, we were not able to extend the parameters to the region in which the triangular OP $m_{\text{Tri}}$ is maximized, as it is still seen to increase at the limiting ratio $J_2/J_1 = -2$.\\

\noindent

As annealing time is increased, the magnitudes of the AFM and triangular OPs are observed to increase, which indicates that the systems are approaching their respective ground states (classical ground state for $m_{\text{AFM}}$ and perturbative one for $m_{\text{Tri}}$). However, we notice that this is not the case for Villain OP $m_{\text{Vil}}$ as it appears to be maximally ordered for an intermediate annealing time which means that this phase is quantum-ordered. In other words, this phase is more sensitive to noise and its thermal state is less ordered than its quantum-quenched state, warranting further study such as the recent \cite{schumm2023}. Now, we would like to quantify the scaling of the OPs with annealing time. \\


\noindent
In Fig. \ref{fig4}a, c, we compute the scaling of $m_{\text{Tri}}$, and $m_{\text{Vil}}$ w.r.t. annealing time for different lattice sizes where we find consistent power-law behavior. Crucially, in Fig. \ref{fig4}a we compare the QA results with matrix product states (MPSs) simulation using the time-dependent variational principle (TDVP) algorithm for time evolution on the $6\times 6$ lattice size. We observe quantitative agreement in this comparison, verifying coherent quantum dynamics (For more details about the MPS-TDVP simulations and convergence analysis, see Methods Section~\ref{MPSmethods}). We also observe in Fig. \ref{fig4}b, d the collapse of the OPs for the different sizes when scaled by the linear system size $L$. While the collapse appears to be better for $m_{\text{Tri}}$, it indicates universal physics independent of the system size for the collapsed instances. Nonetheless, although the MPS simulations are known to be computationally expensive and non-scalable in two dimensions (exponential scaling in the size of the cylinder circumference length\cite{liang94,stoudenmire2012}), scaling on the QA comes with no noticeable overhead, proving the advantage of using QAs as coherent quantum simulators.\\
\begin{figure*}[!htbp]
    \centering
    \includegraphics[scale = 0.25]{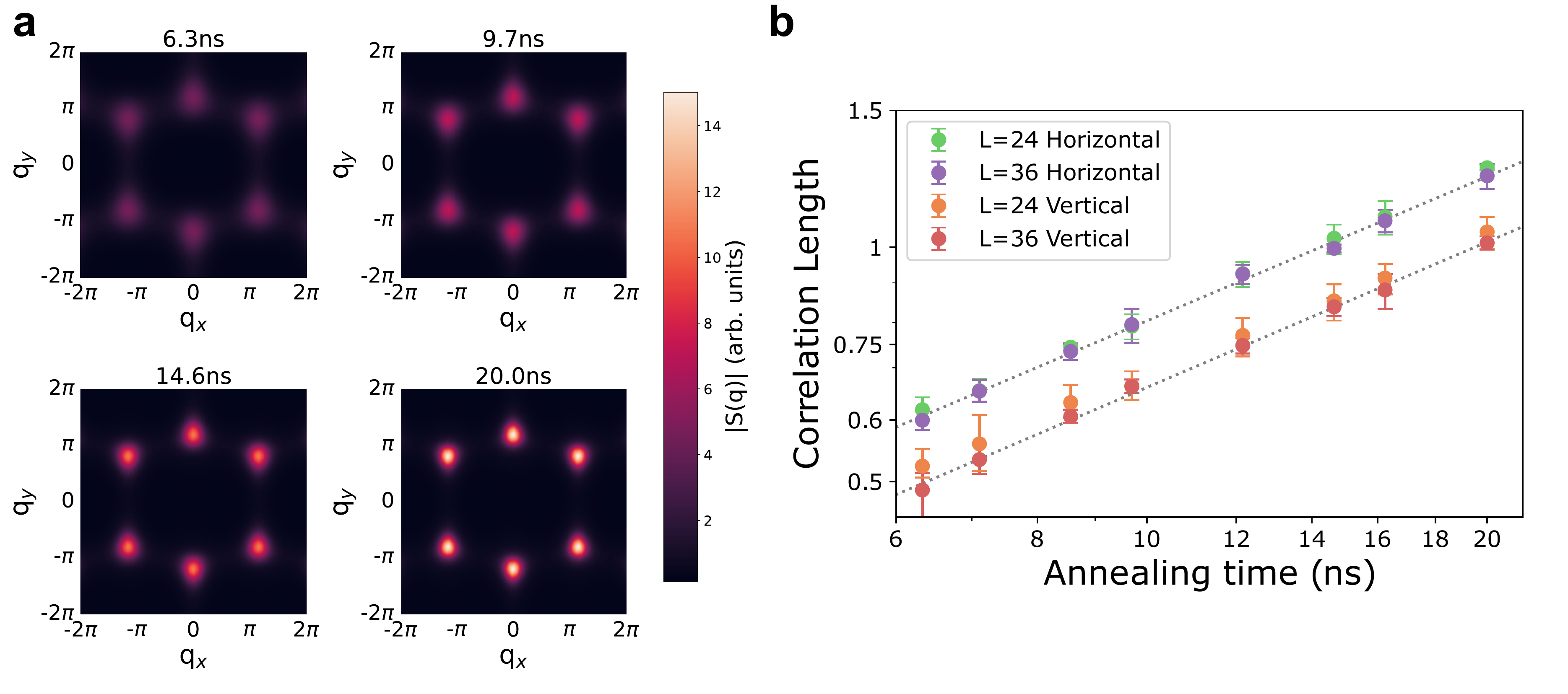}
    \caption{\textbf{Static structure factor (SF) and correlation lengths.} \textbf{a,} The evolution of the SF $S(\textbf{q})$ intensity for different annealing times evaluated on a 36$\times$36 triangular lattice. q$_{x (y)}$ is the momentum along the horizontal (vertical) direction which is open (periodic) with length $L$ ($L/2$). \textbf{b, } Horizontal and vertical Log-Log scaling of the correlation lengths evaluated on the two largest triangular lattice systems $L = 24$ and $36$. Dashed lines correspond to best fits.}
    \label{fig5}
\end{figure*}

\noindent
We report the scaling exponents for the OPs on the largest size $L = 36$ (1296 qubits):
\begin{equation}\label{eq7}
    m_{\text{Tri}} \propto \text{t}_a^{0.35(4)}\text{, }m_{\text{Vil}} \propto \text{t}_a^{0.46(2)}.
\end{equation}
The triangular and Villain models belong to the 3D XY universality class \cite{Moessner2001}, therefore KZM predicts the critical scaling exponent
\begin{equation}\label{eq8}
    m \propto \text{t}_a^{\frac{1-\beta/\nu}{z+1/\nu}} = \text{t}_a^{0.19},
\end{equation}
which does not agree with the observed exponents. To explain this discrepancy, we perform a Monte Carlo (MC) simulation (details in Methods) on the six-state clock model
\begin{equation}\label{eq9}
    \mathcal{H}_{XY} = -  \sum_{\langle i,j\rangle} \text{cos}(\theta_i-\theta_j),
\end{equation}
where $\theta_i = \frac{2\pi q}{6}$, and $q = 0, 1, 2, 3, 4, 5.$ The six-state clock model on a honeycomb lattice (dual triangular lattice) offers an approximate description of the triangular TFIM at weak field \cite{Moessner2001}, i.e. towards the end of the anneal. We find the slope
\begin{equation}\label{eq10}
    m_\text{MC} \propto \text{t}_a^{0.37},
\end{equation}
which is in close agreement with the observed slope on the QA.  This leads us to hypothesize that a coarsening scenario past the quantum critical point dominates the KZM and governs the scaling behavior of the system.\\

\noindent
To investigate this, we look at the scaling of the topological defects (Fig. \ref{fig2}b) on the triangular lattice (Fig. \ref{fig4}e) where we benchmark the lattice size $6\times 6$ using MPS. Our results indicate a scaling exponent
\begin{equation}\label{eq11}
    \rho \propto \text{t}_a^{-1.08(5)},
\end{equation}
again, in disagreement with the KZM prediction
\begin{equation}\label{eq12}
    \rho \propto \text{t}_a^{-\frac{d\nu}{1+z\nu}} = \text{t}_a^{-0.8},
\end{equation}
However, Eq. \ref{eq11} is in close agreement with the coarsening expectation in an XY model \cite{yurke1993} which predicts a scaling with an exponent equal to 1, up to logarithmic corrections. This supports our hypothesis that a coarsening mechanism dominates the system towards the end of the anneal.\\

\noindent
Despite this, looking at Fig. \ref{fig4}f, we find a collapse of the topological defects at different sizes. The behavior of the collapse changes from a power-law for larger system sizes to exponential for the smallest system size $6\times 6$. This indicates a change in the underlying mechanism governing the system dynamics, namely a crossover from KZM to a Landau-Zener regime as the correlation length approaches the system size.  This crossover was also identified in QA results in one-dimensional chains~\cite{King2022}.


\noindent
Finally, we look at the magnetic correlations through the static structure factor (SF)
\begin{equation}\label{eq13}
    S(\textbf{q}) = \frac{1}{N^2}\sum_{i = 1}^N\sum_{j = 1}^N \langle s_is_j\rangle e^{-i\textbf{q}.(\textbf{r}_i - \textbf{r}_j)},
\end{equation}
for the triangular lattice, where $s_i$ denotes the final spin output for site $i$. The SF offers a direct connection between the computational results and experiments on triangular TFIM magnets, such as the rare-earth (RE) heptatantalates $(\text{RETa}_7\text{O}_{19})$ using neutron diffraction and neutron Laue measurements~\cite{arhIsingTriangularlatticeAntiferromagnet2022}. Notably, the 2DXY-KT transition was observed in TFIM compound TmMgGaO$_4$, where the investigation of the melting of the magnetic order was performed as a function of thermal fluctuations (Fig. 5 in~\cite{liKosterlitzThoulessMeltingMagnetic2020}).\\

\noindent
Fig. \ref{fig5}a shows the SF evaluated for the $36\times 36$ lattice at different annealing times up to $\SI{20}{ns}$. We distinguish the six-peak hexagon typical of triangular order. We also observe how the peaks became sharper as the annealing time increased indicating higher order. We obtain the correlation lengths $\xi$ as the inverse FWHM of the peaks and showcase how they scale with annealing time in Fig. \ref{fig5}b for the vertical (periodic, effective length $L/2$) and horizontal (open, length $L$) directions. We observe consistent behavior for the sizes $L = 24$ and $36$ for both directions with the scaling exponent
\begin{equation}\label{eq14}
    \xi \propto \text{t}_a^{0.62(3)},
\end{equation}
as opposed to the KZM expectation
\begin{equation}\label{eq15}
    \xi \propto \text{t}_a^{\frac{\nu}{1+z\nu}} = \text{t}_a^{0.4}.
\end{equation}
But again, our observed scaling is in agreement with the MC result (details in Methods)
\begin{equation}\label{eq16}
    \xi_\text{MC} \propto \text{t}^{0.55},
\end{equation}
further solidifying the coarsening argument.\\

\noindent
Therefore, we conclude that a coarsening mechanism dominates KZM at later stages in the anneal and controls the post-critical dynamics of the systems, as seen through the scaling of the OPs (Eq. \ref{eq7}), topological defects (Eq. \ref{eq11}), and correlation lengths (Eq. \ref{eq14}). Nonetheless, programmable QAs continue to prove their advantage as scalable coherent simulators for quantum systems.

\section{Acknowledgments}

A.A., H.X. and their work with D-Wave is based upon work supported by the Center for Quantum Technologies (CQT) under the Industry-University Cooperative Research Center (IUCRC) Program at the US National Science Foundation (NSF) under Grant No. 2224960. A.A. and A.B. additionally acknowledge funding from Quantum Science Center (QSC), a National Quantum Science Initiative of the Department Of Energy (DOE), managed by Oak Ridge National Laboratory (ORNL) for the MPS effort, and access to Purdue Bell cluster. A.B. thanks D-Wave for access to fast annealing protocols in their devices.
A.N. was supported by Natural Sciences and Engineering Research Council of Canada (NSERC) Alliance Quantum Program (Grant ALLRP-578555), CIFAR and the Canada First Research Excellence Fund, Quantum Materials and Future Technologies Program.  We gratefully acknowledge supporting contributions from technical and non-technical staff at D-Wave.

\bibliography{Main}

\begin{thebibliography}{10}
\expandafter\ifx\csname url\endcsname\relax
  \def\url#1{\burl{#1}}\fi
\expandafter\ifx\csname urlprefix\endcsname\relax\def\urlprefix{URL }\fi
\providecommand{\bibinfo}[2]{#2}
\providecommand{\eprint}[2][]{\url{#2}}
\providecommand{\doi}[1]{\url{https://doi.org/#1}}
\bibcommenthead

\bibitem{SCHMIDT20171}
\bibinfo{author}{Schmidt, B.} \& \bibinfo{author}{Thalmeier, P.}
\newblock \bibinfo{title}{Frustrated two dimensional quantum magnets}.
\newblock \emph{\bibinfo{journal}{Physics Reports}} \textbf{\bibinfo{volume}{703}}, \bibinfo{pages}{1--59} (\bibinfo{year}{2017}).
\newblock \urlprefix\url{https://www.sciencedirect.com/science/article/pii/S0370157317302983}.
\newblock \bibinfo{note}{Frustrated two dimensional quantum magnets}.

\bibitem{Shaginyan_Stephanovich_Msezane_Japaridze_Clark_Amusia_Kirichenko_2019}
\bibinfo{author}{Shaginyan, V.~R.} \emph{et~al.}
\newblock \bibinfo{title}{Theoretical and experimental developments in quantum spin liquid in geometrically frustrated magnets: A review - journal of materials science} (\bibinfo{year}{2019}).
\newblock \urlprefix\url{https://link.springer.com/article/10.1007/s10853-019-04128-w}.

\bibitem{Moessner2001}
\bibinfo{author}{Moessner, R.} \& \bibinfo{author}{Sondhi, S.~L.}
\newblock \bibinfo{title}{{Ising models of quantum frustration}}.
\newblock \emph{\bibinfo{journal}{Physical Review B}} \textbf{\bibinfo{volume}{63}}, \bibinfo{pages}{1--19} (\bibinfo{year}{2001}).

\bibitem{anderson1973}
\bibinfo{author}{Anderson, P.~W.}
\newblock \bibinfo{title}{Resonating valence bonds: A new kind of insulator?}
\newblock \emph{\bibinfo{journal}{Materials Research Bulletin}} \textbf{\bibinfo{volume}{8}}, \bibinfo{pages}{153--160} (\bibinfo{year}{1973}).

\bibitem{Savary2017}
\bibinfo{author}{Savary, L.} \& \bibinfo{author}{Balents, L.}
\newblock \bibinfo{title}{Quantum spin liquids: a review}.
\newblock \emph{\bibinfo{journal}{Reports on Progress in Physics}} \textbf{\bibinfo{volume}{80}}, \bibinfo{pages}{016502} (\bibinfo{year}{2016}).
\newblock \urlprefix\url{https://dx.doi.org/10.1088/0034-4885/80/1/016502}.

\bibitem{DefectReview}
\bibinfo{author}{Teo, J.~C.} \& \bibinfo{author}{Hughes, T.~L.}
\newblock \bibinfo{title}{Topological defects in symmetry-protected topological phases}.
\newblock \emph{\bibinfo{journal}{Annual Review of Condensed Matter Physics}} \textbf{\bibinfo{volume}{8}}, \bibinfo{pages}{211--237} (\bibinfo{year}{2017}).
\newblock \urlprefix\url{https://doi.org/10.1146/annurev-conmatphys-031016-025154}.

\bibitem{Blankschtein1984a}
\bibinfo{author}{Blankschtein, D.}, \bibinfo{author}{Ma, M.}, \bibinfo{author}{Berker, A.~N.}, \bibinfo{author}{Grest, G.~S.} \& \bibinfo{author}{Soukoulis, C.~M.}
\newblock \bibinfo{title}{{Orderings of a stacked frustrated triangular system in three dimensions}}.
\newblock \emph{\bibinfo{journal}{Physical Review B}} \textbf{\bibinfo{volume}{29}}, \bibinfo{pages}{5250--5252} (\bibinfo{year}{1984}).
\newblock \urlprefix\url{https://link.aps.org/doi/10.1103/PhysRevB.29.5250}.

\bibitem{Blankschtein1984}
\bibinfo{author}{Blankschtein, D.}, \bibinfo{author}{Ma, M.} \& \bibinfo{author}{Berker, A.~N.}
\newblock \bibinfo{title}{{Fully and partially frustrated simple-cubic Ising models: Landau-Ginzburg-Wilson theory}}.
\newblock \emph{\bibinfo{journal}{Physical Review B}} \textbf{\bibinfo{volume}{30}}, \bibinfo{pages}{1362--1365} (\bibinfo{year}{1984}).
\newblock \urlprefix\url{https://link.aps.org/doi/10.1103/PhysRevB.30.1362}.

\bibitem{sachdev1999}
\bibinfo{author}{Sachdev, S.}
\newblock \bibinfo{title}{Quantum phase transitions}.
\newblock \emph{\bibinfo{journal}{Physics world}} \textbf{\bibinfo{volume}{12}}, \bibinfo{pages}{33} (\bibinfo{year}{1999}).

\bibitem{Jalabert1991}
\bibinfo{author}{Jalabert, R.~A.} \& \bibinfo{author}{Sachdev, S.}
\newblock \bibinfo{title}{{Spontaneous alignment of frustrated bonds in an anisotropic, three-dimensional Ising model}}.
\newblock \emph{\bibinfo{journal}{Physical Review B}} \textbf{\bibinfo{volume}{44}}, \bibinfo{pages}{686--690} (\bibinfo{year}{1991}).
\newblock \urlprefix\url{https://link.aps.org/doi/10.1103/PhysRevB.44.686}.

\bibitem{Moessner2000}
\bibinfo{author}{Moessner, R.}, \bibinfo{author}{Sondhi, S.~L.} \& \bibinfo{author}{Chandra, P.}
\newblock \bibinfo{title}{{Two-Dimensional Periodic Frustrated Ising Models in a Transverse Field}}.
\newblock \emph{\bibinfo{journal}{Physical Review Letters}} \textbf{\bibinfo{volume}{84}}, \bibinfo{pages}{4457--4460} (\bibinfo{year}{2000}).
\newblock \urlprefix\url{https://link.aps.org/doi/10.1103/PhysRevLett.84.4457}.

\bibitem{Isakov2003}
\bibinfo{author}{Isakov, S.~V.} \& \bibinfo{author}{Moessner, R.}
\newblock \bibinfo{title}{{Interplay of quantum and thermal fluctuations in a frustrated magnet}}.
\newblock \emph{\bibinfo{journal}{Physical Review B}} \textbf{\bibinfo{volume}{68}}, \bibinfo{pages}{104409} (\bibinfo{year}{2003}).
\newblock \urlprefix\url{https://link.aps.org/doi/10.1103/PhysRevB.68.104409}.

\bibitem{villain1977}
\bibinfo{author}{Villain, J.}
\newblock \bibinfo{title}{Spin glass with non-random interactions}.
\newblock \emph{\bibinfo{journal}{Journal of Physics C: Solid State Physics}} \textbf{\bibinfo{volume}{10}}, \bibinfo{pages}{1717} (\bibinfo{year}{1977}).

\bibitem{King2018}
\bibinfo{author}{King, A.~D.} \emph{et~al.}
\newblock \bibinfo{title}{{Observation of topological phenomena in a programmable lattice of 1,800 qubits}}.
\newblock \emph{\bibinfo{journal}{Nature}} \textbf{\bibinfo{volume}{560}}, \bibinfo{pages}{456--460} (\bibinfo{year}{2018}).
\newblock \urlprefix\url{https://www.nature.com/articles/s41586-018-0410-x}.

\bibitem{kosterlitz1973}
\bibinfo{author}{Kosterlitz, J.} \& \bibinfo{author}{Thouless, D.}
\newblock \bibinfo{title}{Ordering, metastability and phase transitions in two-dimensional systems 1973}.
\newblock \emph{\bibinfo{journal}{J. Phys.: Condens. Matter}} \textbf{\bibinfo{volume}{6}}, \bibinfo{pages}{1181} (\bibinfo{year}{1973}).

\bibitem{King2021}
\bibinfo{author}{King, A.~D.} \emph{et~al.}
\newblock \bibinfo{title}{{Scaling advantage over path-integral Monte Carlo in quantum simulation of geometrically frustrated magnets}}.
\newblock \emph{\bibinfo{journal}{Nature Communications}} \textbf{\bibinfo{volume}{12}}, \bibinfo{pages}{1113} (\bibinfo{year}{2021}).
\newblock \urlprefix\url{https://www.nature.com/articles/s41467-021-20901-5 http://www.nature.com/articles/s41467-021-20901-5}.

\bibitem{bauza2024scaling}
\bibinfo{author}{Bauza, H.~M.} \& \bibinfo{author}{Lidar, D.~A.}
\newblock \bibinfo{title}{Scaling advantage in approximate optimization with quantum annealing} (\bibinfo{year}{2024}).
\newblock \eprint{2401.07184}.

\bibitem{King2022}
\bibinfo{author}{King, A.~D.} \emph{et~al.}
\newblock \bibinfo{title}{{Coherent quantum annealing in a programmable 2,000 qubit Ising chain}}.
\newblock \emph{\bibinfo{journal}{Nature Physics}} \textbf{\bibinfo{volume}{18}}, \bibinfo{pages}{1324--1328} (\bibinfo{year}{2022}).
\newblock \urlprefix\url{http://arxiv.org/abs/2202.05847 https://www.nature.com/articles/s41567-022-01741-6}.

\bibitem{kibble1976}
\bibinfo{author}{Kibble, T.~W.}
\newblock \bibinfo{title}{Topology of cosmic domains and strings}.
\newblock \emph{\bibinfo{journal}{Journal of Physics A: Mathematical and General}} \textbf{\bibinfo{volume}{9}}, \bibinfo{pages}{1387} (\bibinfo{year}{1976}).

\bibitem{zurek1985}
\bibinfo{author}{Zurek, W.~H.}
\newblock \bibinfo{title}{Cosmological experiments in superfluid helium?}
\newblock \emph{\bibinfo{journal}{Nature}} \textbf{\bibinfo{volume}{317}}, \bibinfo{pages}{505--508} (\bibinfo{year}{1985}).

\bibitem{Keesling2019}
\bibinfo{author}{Keesling, A.} \emph{et~al.}
\newblock \bibinfo{title}{Quantum kibble–zurek mechanism and critical dynamics on a programmable rydberg simulator} (\bibinfo{year}{2019}).
\newblock \urlprefix\url{http://dx.doi.org/10.1038/s41586-019-1070-1}.

\bibitem{Ebadi2021}
\bibinfo{author}{Ebadi, S.} \emph{et~al.}
\newblock \bibinfo{title}{Quantum phases of matter on a 256-atom programmable quantum simulator} (\bibinfo{year}{2021}).
\newblock \urlprefix\url{http://dx.doi.org/10.1038/s41586-021-03582-4}.

\bibitem{King2023}
\bibinfo{author}{King, A.~D.} \emph{et~al.}
\newblock \bibinfo{title}{{Quantum critical dynamics in a 5000-qubit programmable spin glass}}.
\newblock \emph{\bibinfo{journal}{Nature}} \bibinfo{pages}{1--41} (\bibinfo{year}{2022}).
\newblock \urlprefix\url{http://arxiv.org/abs/2207.13800}.

\bibitem{schumm2023}
\bibinfo{author}{Schumm, G.}, \bibinfo{author}{Shao, H.}, \bibinfo{author}{Guo, W.}, \bibinfo{author}{Mila, F.} \& \bibinfo{author}{Sandvik, A.~W.}
\newblock \bibinfo{title}{Primary and secondary order parameters in the fully frustrated transverse field ising model on the square lattice} (\bibinfo{year}{2023}).
\newblock \eprint{2309.02407}.

\bibitem{liang94}
\bibinfo{author}{Liang, S.} \& \bibinfo{author}{Pang, H.}
\newblock \bibinfo{title}{Approximate diagonalization using the density matrix renormalization-group method: A two-dimensional-systems perspective}.
\newblock \emph{\bibinfo{journal}{Phys. Rev. B}} \textbf{\bibinfo{volume}{49}}, \bibinfo{pages}{9214--9217} (\bibinfo{year}{1994}).
\newblock \urlprefix\url{https://link.aps.org/doi/10.1103/PhysRevB.49.9214}.

\bibitem{stoudenmire2012}
\bibinfo{author}{Stoudenmire, E.~M.} \& \bibinfo{author}{White, S.~R.}
\newblock \bibinfo{title}{Studying two-dimensional systems with the density matrix renormalization group}.
\newblock \emph{\bibinfo{journal}{Annu. Rev. Condens. Matter Phys.}} \textbf{\bibinfo{volume}{3}}, \bibinfo{pages}{111--128} (\bibinfo{year}{2012}).

\bibitem{yurke1993}
\bibinfo{author}{Yurke, B.}, \bibinfo{author}{Pargellis, A.~N.}, \bibinfo{author}{Kovacs, T.} \& \bibinfo{author}{Huse, D.~A.}
\newblock \bibinfo{title}{Coarsening dynamics of the xy model}.
\newblock \emph{\bibinfo{journal}{Phys. Rev. E}} \textbf{\bibinfo{volume}{47}}, \bibinfo{pages}{1525--1530} (\bibinfo{year}{1993}).
\newblock \urlprefix\url{https://link.aps.org/doi/10.1103/PhysRevE.47.1525}.

\bibitem{arhIsingTriangularlatticeAntiferromagnet2022}
\bibinfo{author}{Arh, T.} \emph{et~al.}
\newblock \bibinfo{title}{The {{Ising}} triangular-lattice antiferromagnet neodymium heptatantalate as a quantum spin liquid candidate} \textbf{\bibinfo{volume}{21}}, \bibinfo{pages}{416--422}.
\newblock \urlprefix\url{https://www.nature.com/articles/s41563-021-01169-y}.

\bibitem{liKosterlitzThoulessMeltingMagnetic2020}
\bibinfo{author}{Li, H.} \emph{et~al.}
\newblock \bibinfo{title}{Kosterlitz-{{Thouless}} melting of magnetic order in the triangular quantum {{Ising}} material {{TmMgGaO4}}} \textbf{\bibinfo{volume}{11}}, \bibinfo{pages}{1111}.
\newblock \urlprefix\url{https://doi.org/10.1038/s41467-020-14907-8}.

\bibitem{chern_tutorial_2023}
\bibinfo{author}{Chern, K.}, \bibinfo{author}{Boothby, K.}, \bibinfo{author}{Raymond, J.}, \bibinfo{author}{Farré, P.} \& \bibinfo{author}{King, A.~D.}
\newblock \bibinfo{title}{Tutorial: Calibration refinement in quantum annealing} (\bibinfo{year}{2023}).
\newblock \eprint{2304.10352}.

\bibitem{ITensor}
\bibinfo{author}{Fishman, M.}, \bibinfo{author}{White, S.~R.} \& \bibinfo{author}{Stoudenmire, E.~M.}
\newblock \bibinfo{title}{{The ITensor Software Library for Tensor Network Calculations}}.
\newblock \emph{\bibinfo{journal}{SciPost Phys. Codebases}} \bibinfo{pages}{4} (\bibinfo{year}{2022}).
\newblock \urlprefix\url{https://scipost.org/10.21468/SciPostPhysCodeb.4}.

\bibitem{ITensor-r0.3}
\bibinfo{author}{Fishman, M.}, \bibinfo{author}{White, S.~R.} \& \bibinfo{author}{Stoudenmire, E.~M.}
\newblock \bibinfo{title}{{Codebase release 0.3 for ITensor}}.
\newblock \emph{\bibinfo{journal}{SciPost Phys. Codebases}} \bibinfo{pages}{4--r0.3} (\bibinfo{year}{2022}).
\newblock \urlprefix\url{https://scipost.org/10.21468/SciPostPhysCodeb.4-r0.3}.

\bibitem{Zhu2015}
\bibinfo{author}{Zhu, Z.} \& \bibinfo{author}{White, S.~R.}
\newblock \bibinfo{title}{Spin liquid phase of the $s=\frac{1}{2}\phantom{\rule{4.pt}{0ex}}{J}_{1}\ensuremath{-}{J}_{2}$ heisenberg model on the triangular lattice}.
\newblock \emph{\bibinfo{journal}{Phys. Rev. B}} \textbf{\bibinfo{volume}{92}}, \bibinfo{pages}{041105} (\bibinfo{year}{2015}).
\newblock \urlprefix\url{https://link.aps.org/doi/10.1103/PhysRevB.92.041105}.

\bibitem{TDVP}
\bibinfo{author}{Haegeman, J.}, \bibinfo{author}{Lubich, C.}, \bibinfo{author}{Oseledets, I.}, \bibinfo{author}{Vandereycken, B.} \& \bibinfo{author}{Verstraete, F.}
\newblock \bibinfo{title}{Unifying time evolution and optimization with matrix product states}.
\newblock \emph{\bibinfo{journal}{Phys. Rev. B}} \textbf{\bibinfo{volume}{94}}, \bibinfo{pages}{165116} (\bibinfo{year}{2016}).
\newblock \urlprefix\url{https://link.aps.org/doi/10.1103/PhysRevB.94.165116}.

\bibitem{Zaletel2015}
\bibinfo{author}{Zaletel, M.~P.}, \bibinfo{author}{Mong, R. S.~K.}, \bibinfo{author}{Karrasch, C.}, \bibinfo{author}{Moore, J.~E.} \& \bibinfo{author}{Pollmann, F.}
\newblock \bibinfo{title}{Time-evolving a matrix product state with long-ranged interactions}.
\newblock \emph{\bibinfo{journal}{Phys. Rev. B}} \textbf{\bibinfo{volume}{91}}, \bibinfo{pages}{165112} (\bibinfo{year}{2015}).
\newblock \urlprefix\url{https://link.aps.org/doi/10.1103/PhysRevB.91.165112}.

\bibitem{newvilleLMFITNonLinearLeastSquare2014}
\bibinfo{author}{Newville, M.}, \bibinfo{author}{Stensitzki, T.}, \bibinfo{author}{Allen, D.~B.} \& \bibinfo{author}{Ingargiola, A.}
\newblock \bibinfo{title}{{{LMFIT}}: {{Non-Linear Least-Square Minimization}} and {{Curve-Fitting}} for {{Python}}}.
\newblock \bibinfo{howpublished}{Zenodo} (\bibinfo{year}{2014}).

\end{thebibliography}

\section{Methods}\label{sec2}
\subsection{Quantum Annealing}
All quantum annealing data were taken from D-Wave Advantage 4.1 QA. The annealing schedule is depicted in Extended Data Fig. \ref{figSch}. We performed an iterative calibration refinement method, called shimming, to calibrate the qubits. The detailed shimming routines are explained in the next subsection. Each shimming iteration took 100 samples which may consist of multiple disjoint lattices depending on the lattice size (Extended Data Table \ref{tab:qa_detail}). We performed 1500 shimming iterations for each lattice size at every anneal time and collected several samples at the end after the shimming parameters stabilized. The specific numbers of disjoint lattices per sample and samples collected for computations are shown in Table \ref{tab:qa_detail}.

\subsection{Shimming}\label{sec:shimming}
\noindent
Frustrated systems are extremely sensitive to noise perturbations stemming from the fact that these perturbations break their vast degeneracy. For this reason, following \cite{King2022}, \cite{chern_tutorial_2023} we perform three hardware calibration measures:
\begin{itemize}
    \item Flux Bias Shim
    \item Couplers' Strength Shim
    \item Anneal Offset Shim
\end{itemize}
which we now explain. For shimming plots refer to Extended Data Fig. \ref{figShimming} and \ref{figShimStats}. For lattice sizes with multiple copies of disjoint lattices, shimming is performed based on their combined statistics.\\

\noindent
\textit{Flux Bias Shim}\\

\noindent
The Hamiltonian, Eq. \ref{eq2}, has zero longitudinal magnetic field. Therefore, the expected magnetization of each qubit should be zero after measuring a statistical number of samples. We iteratively adjust the flux bias for each qubit where at the $k$-th iteration, with the average magnetization for qubit $i$ of the previous iteration being $\langle m \rangle ^k_i$, we set the flux bias to
\begin{equation}\label{eq:shimming_fb}
    \phi_i^k = \phi_{i}^{k-1} - \delta_{\phi}  \langle m \rangle ^k_i,
\end{equation}
$\delta_{\phi} = 2e-6$. This flux bias corrects the behavior of the qubits and gets rid of any internal bias to point in a specific direction.\\

\noindent
\textit{Couplers' Strength Shim}\\

\noindent
Given a coupler $J_{ij}$, its probability to be frustrated is 
\begin{equation}
    f_{ij} = (\text{sign}(J_{ij}) \langle m_i m_j \rangle + 1) / 2.
\end{equation}
The symmetries present in the problem will give some expectations of what the average frustration should be for the different couplers. For example, a 1D FM ring is rotationally invariant, thus all the couplers should have the same frustration probability and are said to be in the same orbit. For our problem, the couplers are arranged in a cylinder with alternating FM and AFM bonds along the periodic direction (Fig. \ref{fig1}c). Therefore, every other coupler in the same column belongs to the same orbit.\\
The orbit that contains $f_{ij}$ has the average frustration probability:
\begin{equation}
    f_{O_{ij}} = \sum_{i^{'}j^{'} \in O_{ij}} \frac{f_{i^{'}j^{'}}}{|O_{ij}|},
\end{equation}
where $O_{ij}$ is the orbit containing the coupler $f_{ij}$, and $|O_{ij}|$ is its size.
At the $k$-th iteration, we adjust the coupler strength as
\begin{equation}
    J_{ij}^k = J_{ij}^{k-1} + \text{sign}(J_{ij})  \delta_{f}  (f_{ij} - f_{O_{ij}}),
\end{equation}
$\delta_{f} = 2.5e-3$. This procedure ensures that all symmetrically equivalent couplers are acting with the same effective strength.\\

\noindent
\textit{Anneal Offset Shim}\\

\noindent
The qubits and couplers of Advantage 4.1 QA are arranged on 8 different annealing lines that perform the annealing independently from $s = 0$ to $s = 1$. For the extremely fast anneals we are performing, the 8 annealing lines may not be in sync with discrepancies ranging from a fraction of a nanosecond to one nanosecond. To correct for this, we obtain the annealing offsets from a 1D FM ring where we enforce the average frustration per annealing line to be equal, as was done in \cite{King2022}.

\subsection{MPS Simulation}\label{MPSmethods}
\noindent
In this work, MPS simulations were performed using the ITensor Julia library \cite{ITensor, ITensor-r0.3}. The triangular lattice was simulated using open-ended cylinders, with the axis along
the x direction and one of the three bond directions aligned along the y axis, realizing a YC-6 cylinder~\cite{Zhu2015} (see Fig. \ref{fig1}a).  Time evolution was performed using the time-dependent variational principle (TDVP) algorithm \cite{TDVP} with a two-site update where a convergence for the computed physical quantities was observed at a bond dimension $\chi = 150$. Care was taken (by playing with the size of the time step $dt$) to ensure that the bond dimension grows with time since the algorithm starts from the product state $\prod_i \ket{\rightarrow}_i$. We also compared different methods to simulate the anneal dynamics. Extended Data Fig~\ref{figMPSCompare} shows comparison analysis as a function of the time step size $dt$ for the TDVP method against the $W_I$ and $W_{II}$ approximations introduced in Ref.~\cite{Zaletel2015}. Clearly, the ability to use large time steps makes TDVP the best choice in terms of speed and accuracy.
The initial state is the Bell state (Eq. \ref{eq3}), and the Hamiltonian parameters $\Gamma(s)$ and $\mathcal{J}(s)$ followed the exact annealing schedule of Advantage 4.1. The relationship between the simulation time $\text{t}_{sim}$ and the annealing time in ns $\text{t}_a$ is $\text{t}_{sim} = \pi\text{t}_a$, and a time step of $dt = 0.05$ was used. The relevant physical quantities studied in the main text (OPs and structure factors) are then computed 
by sampling the wave-function at the end of the annealing process which collapses it into the computational basis. We do not simply obtain expectation values which could be deceptive for symmetry unbroken superposition states obtained in simulations as opposed to symmetry broken states defining our OPs. For example, a quantum ferromagnetic ground state $\ket{\uparrow \uparrow \uparrow ..}+\ket{\downarrow \downarrow \downarrow ..}$ has $m = \sum_i\langle \sigma_i^z \rangle = 0$, but measuring and collapsing the state correctly yields $|m| = 1$.

\subsection{Statistical Methods}
Error bars from Fig. \ref{fig4} are generated through a bootstrap method where we first average the observables from each QPU call of 100 samples, and then compute the standard deviation out of these averages. To compute confidence intervals for correlation lengths in Fig. \ref{fig5}, we used \textit{lmfit}'s \cite{newvilleLMFITNonLinearLeastSquare2014} \textit{conf\_interval} method. It employs F-test in Eq. \ref{eq:ftest} to compare $\chi^2$ statistics of the null model with our best-found fitting parameters to an alternate model where one of the parameters is fixed. 
\begin{equation} \label{eq:ftest}
    F(P_{fix}, N-P) = (\frac{\chi^2_f}{\chi^2_0} - 1)\frac{N-P}{P_{fix}}
\end{equation}
Here $N$ is the number of data points and $P$ is the number of parameters of the null model. $P_{fix}$ is the number of fixed parameters.\\
To obtain the best-fit slopes, we bootstrap by resampling the data points of the largest system size ($36 \times 36$) 200 times. We then report the mean and the 95\% confidence interval of the bootstrap.

\subsection{Monte Carlo Coarsening Simulation}

Here we compare dynamic scaling of the order parameter and correlation length with a quasi-classical coarsening scenario, as hypothesized to dominate scaling due to postcritical dynamics.  In the perturbative regime ($\Gamma \ll \mathcal J$), the effective model is a two-dimensional six-state ``clock'' XY model~\cite{Isakov2003,King2018}.  The XY spins lie on the plaquettes of the triangular lattice, i.e., on the sites of the dual honeycomb lattice.\\
\noindent
To simulate a coarsening dynamics we proceed with an absorbing Markov-chain Monte Carlo (MCMC) process with both two-site and one-site updates, to avoid local minima.  The energy of the state $\bm{\theta} = \{ \theta_1, \ldots, \theta_N\}$ is calculated as 
\begin{equation}
\mathcal{H}_{XY} = -\sum_{\langle i,j\rangle}\cos{(\theta_i-\theta_j}).
\end{equation}
At each time step, we first process all edges of the model in random order, proposing new angles uniformly at random for both endpoints, and accepting if and only if the energy is not increased.  We then process all individual spins in random order, proposing new angles one at a time.\\
\noindent
This was performed on a honeycomb lattice with $L=120$ and fully periodic boundary conditions, constructed from a square lattice by deletion of edges.  100 independent replicas were run for 1000 time steps.  We computed $\langle m\rangle$ as the average rotor value (unit vectors with angle $\theta_i$) and computed the correlation between two spins as $\cos{(\theta_i-\theta_j)}$ (Extended Data Fig. \ref{figMC}).  Correlation lengths were computed based on distances 5 to 20.

\subsection{Correlation Length Extraction}
Correlation lengths were extracted from the peaks of the structure factor (Eq. \ref{eq13}) by fitting to the Pseudo-Voigt function
\begin{equation}
    V(x) = \eta G(x) + (1-\eta)L(x), \text{ } 0\leq\eta\leq1,
\end{equation}
where $G(x)$ is the Gaussian function $\frac{1}{\sqrt{2\pi}\sigma}e^{-\frac{(x-x_0)^2}{2\sigma^2}}, \text{ }\sigma = \frac{\Gamma}{2\sqrt{2\ln{2}}}$ and $L(x)$ is the Lorenzian function $\frac{1}{\pi}.\frac{\Gamma /2}{(x-x_0)^2+(\Gamma /2)^2}$. Here $x_0$ is the maximum momentum position and $\Gamma$ is the full-width at half maximum (FWHM) from which the correlation length follows as $\xi = 1/\Gamma$.

\backmatter



\begin{appendices}
\newpage
\onecolumn
\section{\centering{Extended Data: Quantum Quench Dynamics of Geometrically Frustrated Ising Models}}\label{secA1}

\begin{figure*}[!htbp]
    \centering
    \hspace*{-1cm}
    \includegraphics[scale = 0.5]{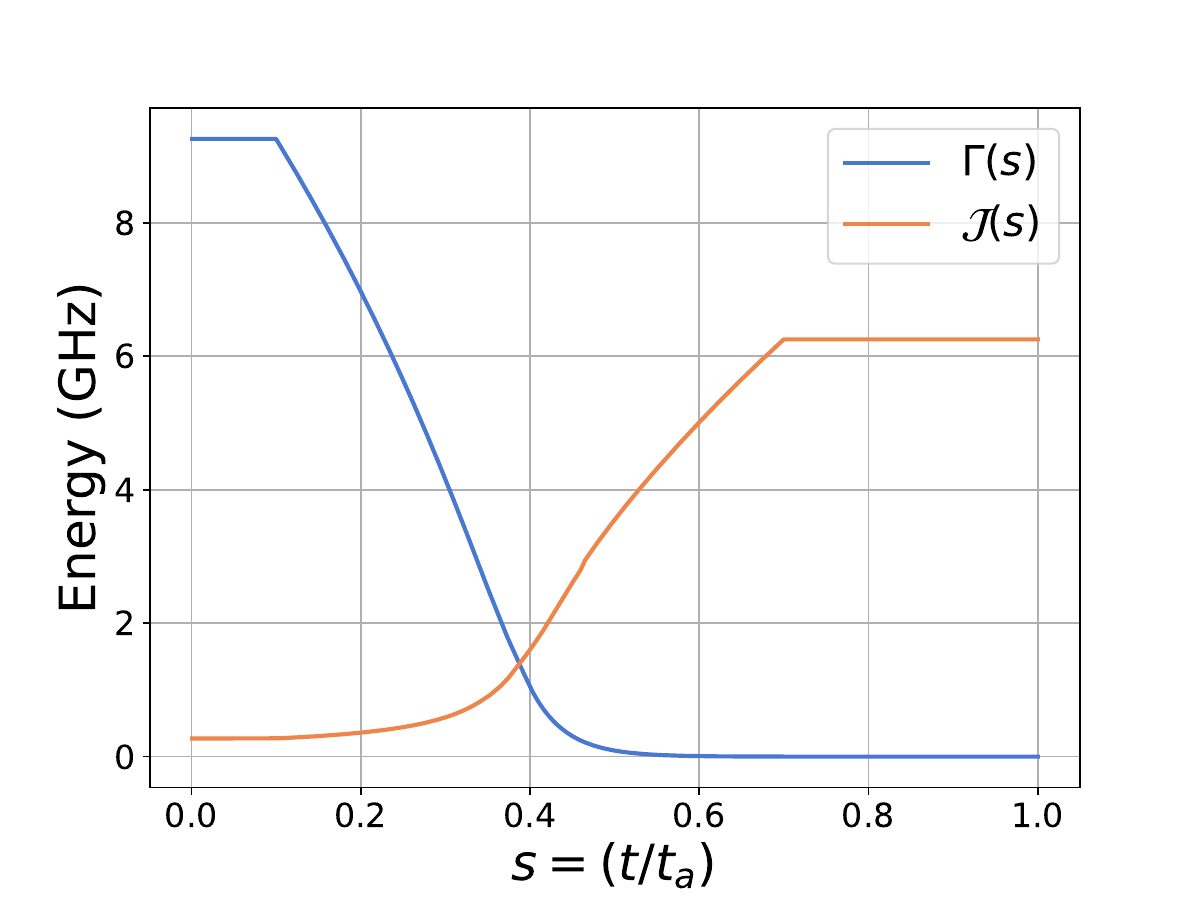}
    \caption{\textbf{Annealing schedule of Advatnage 4.1 QA.} Evolution of the Hamiltonian parameters (Eq. \ref{eq2}) $\Gamma(s)$ and $\mathcal{J}(s)$ as a function of the normalized annealing time $s = \frac{\text{t}}{\text{t}_a}$.} 
    \label{figSch}
\end{figure*}

\begin{table}[b]
\resizebox{.7\textwidth}{!}{
\begin{tabular}{lll}
\hline
Lattice Size & Lattices per Sample& Collected Samples  \\
\hline
$6\times 6$ & 46 & 2200 \\
$12\times 12$ & 18 &  5600  \\
$24\times 24$ & 2 &  25000  \\
$36\times 36$ & 1 & 50000\\
\hline
\end{tabular}
}
\caption{The numbers of disjoint lattices per sample and numbers of samples collected for computations in triangular lattice experiments.}\label{tab:qa_detail}
\end{table}

\begin{figure*}[!htbp]
    \centering
    \hspace*{-1.4cm}
    \includegraphics[scale = 0.42]{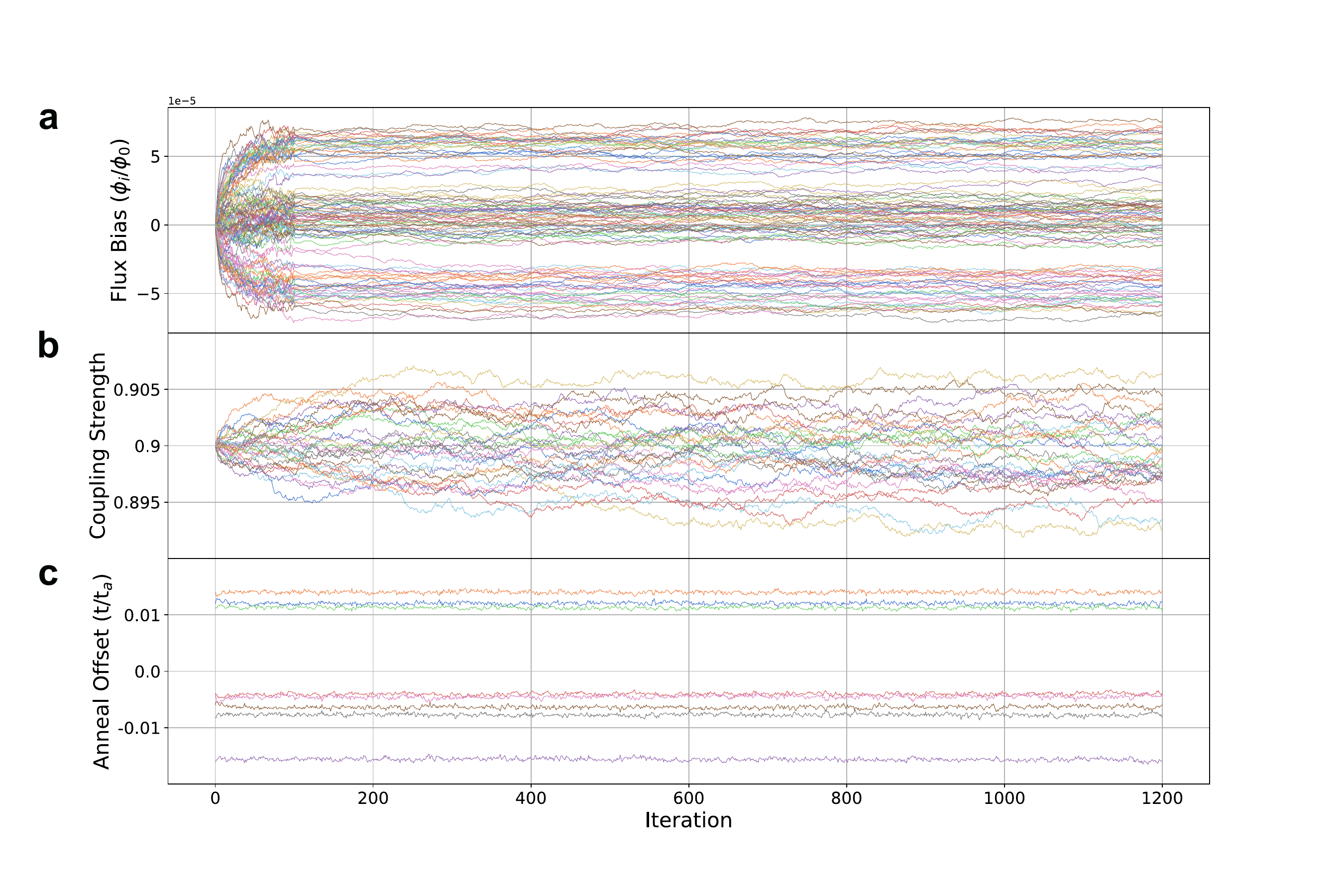}
    \caption{\textbf{Shimming evolution plots at $\text{t}_a = \SI{20}{ns}$.} \textbf{a,} Flux bias strength evolution for 100 randomly sampled qubits. \textbf{b, } Evolution of the coupling strength of 30 random couplers where the initial coupling strength is 0.9. \textbf{c, } The anneal offset of the 8 different annealing lines, initialized from a 1D chain.}
    \label{figShimming}
\end{figure*}

\begin{figure*}[!htbp]
    \centering
    \hspace*{-1cm}
    \includegraphics[scale = 0.24]{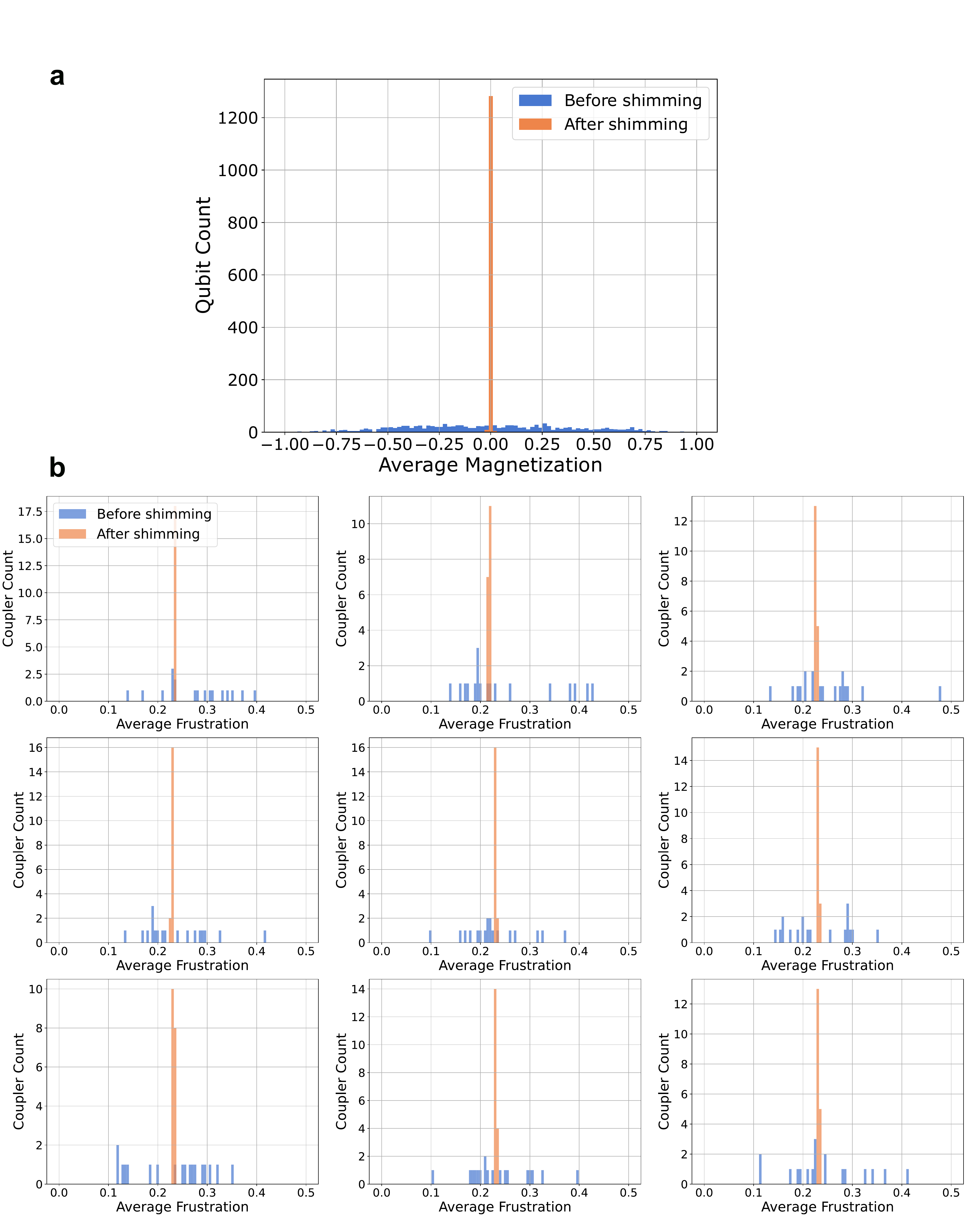}
    \caption{\textbf{Shimming statistics for a $36 \times 36$ lattice at $\text{t}_a = \SI{20}{ns}$.} \textbf{a,} Average qubit magnetization before and after shimming where it becomes peaked at 0 which is the expected value in the absence of a longitudinal magnetic field. \textbf{b, } A 9 orbit sample (out of 106 orbits) of the average coupler frustration before and after shimming. Couplers are seen to converge to a common average indicating that they act with the same effective strength.}
    \label{figShimStats}
\end{figure*}

\clearpage

\begin{figure*}[!htbp]
    \centering
    \hspace*{-2.5cm}
    \includegraphics[scale = 0.27]{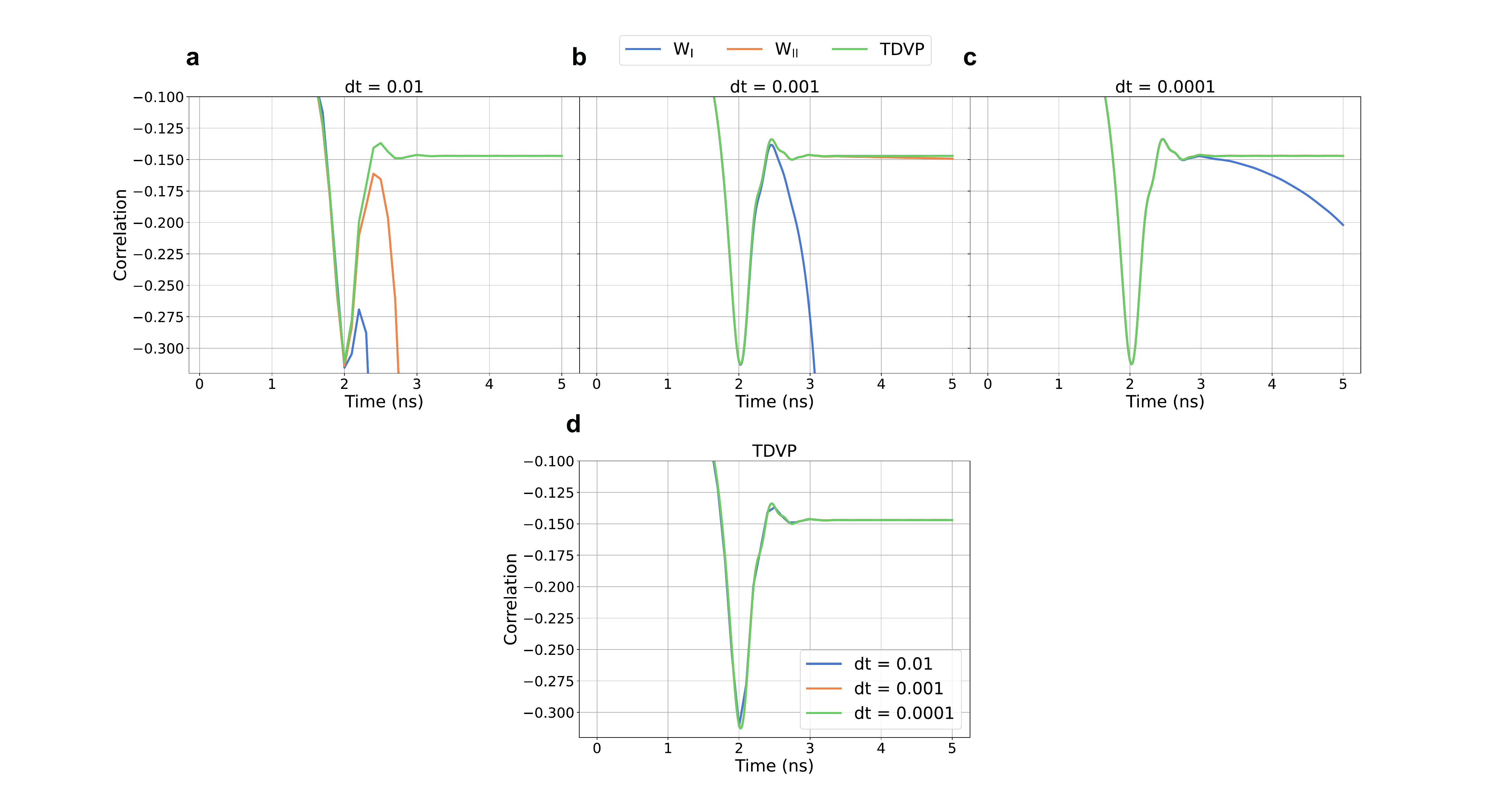}
    \caption{\textbf{Comparison between W$_{\text{I}}$, W$_{\text{II}}$, and TDVP time evolution algorithms for a 4x4 lattice instance.} \textbf{a, b, c,} Correlation $\langle \sigma_z \sigma_z\rangle$ between two bulk qubits as a function of time with a fixed bond dimension $\chi$ = 32 for time steps dt = 0.01, 0.001, 0.0001. TDVP (with two-site update) is observed to converge for a larger time step, thus proving to be the more efficient method to simulate 2D MPS dynamics. \textbf{d,} TDVP correlation for the different time steps. }
    \label{figMPSCompare}
\end{figure*}

\begin{figure*}[!htbp]
    \centering
    \includegraphics[scale = 0.5]{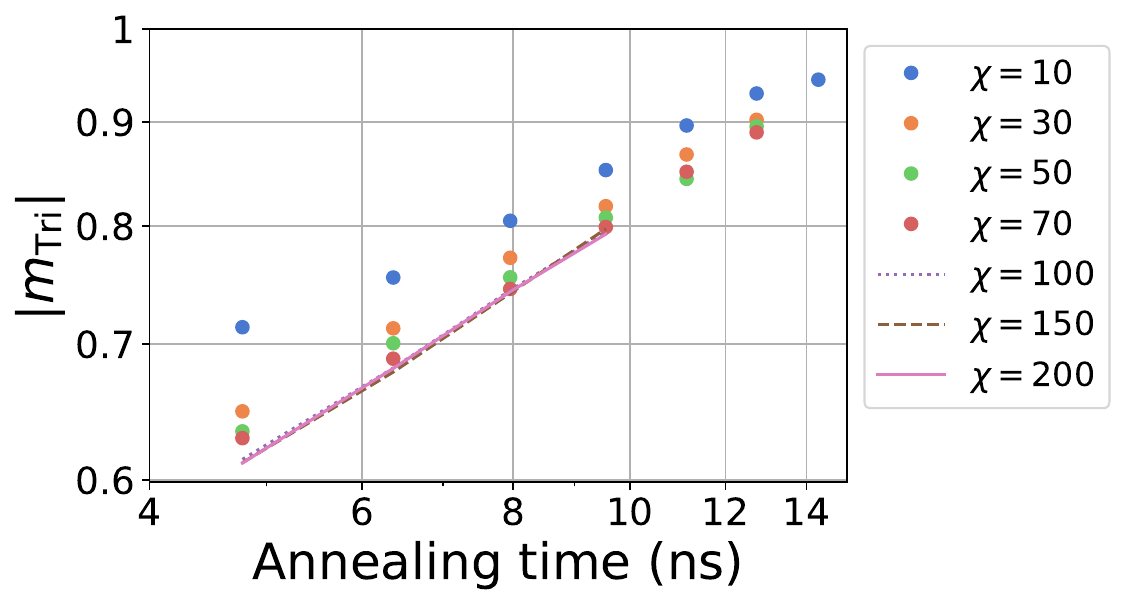}
    \caption{\textbf{$6\times 6$ triangular lattice OP convergence with TDVP time evolution.} Convergence of the triangular lattice OP with respect to the maximum TDVP bond dimension $\chi$ at different annealing times. A time step of $dt=0.05$ was used for simulation times equal to $\pi \times$ annealing time.}
    \label{figConvergence}
\end{figure*}

\clearpage

\begin{figure*}[!htbp]
    \centering
    \hspace*{-0.5cm}
    \includegraphics[scale = 0.5]{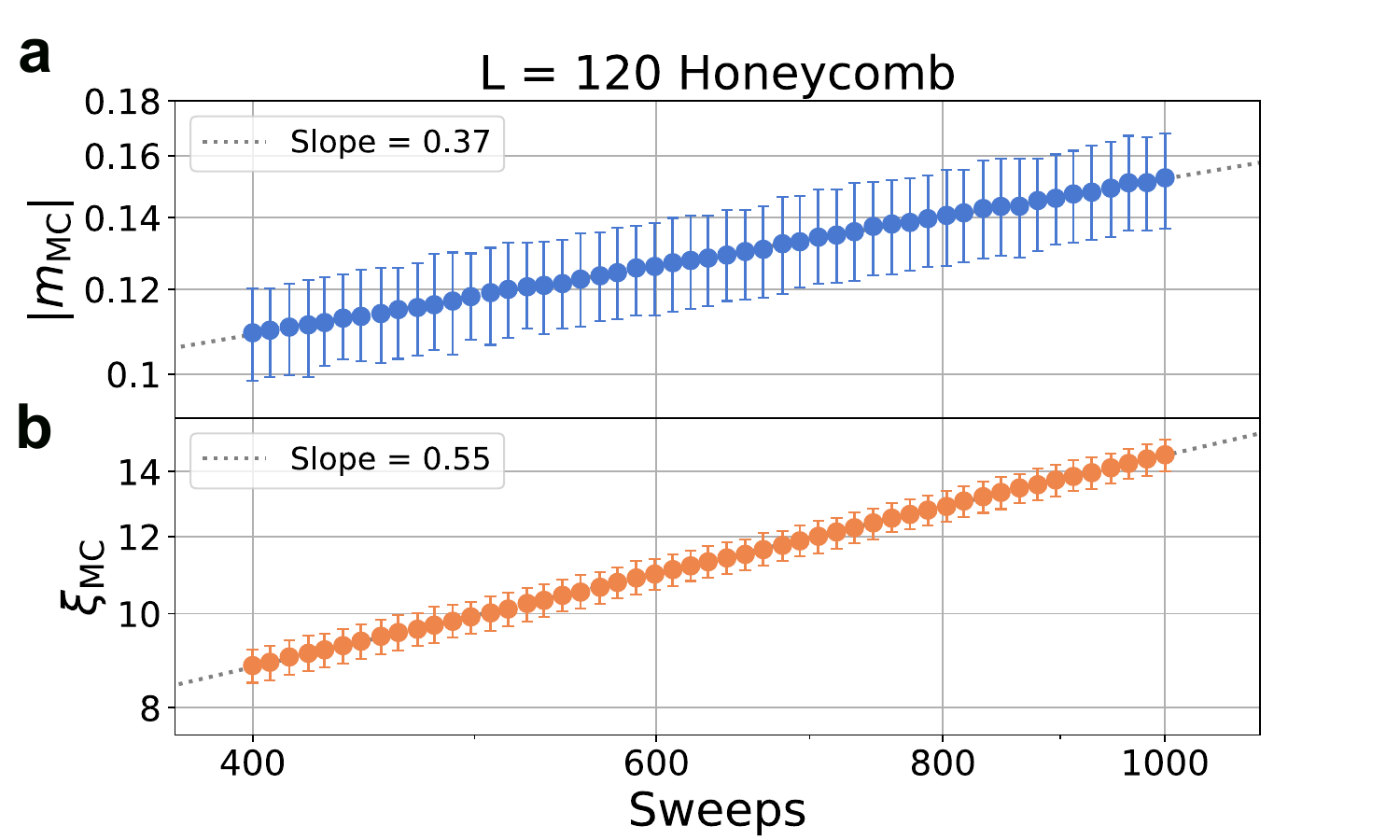}
    \caption{\textbf{Markov-chain Monte Carlo scaling of the six-state clock model on a honeycomb lattice.} \textbf{a,} Order parameter scaling. \textbf{b, } Correlation length scaling. Error bars are bootstrap 95\% C.I. over 100 replicas.}
    \label{figMC}
\end{figure*}
\clearpage
\begin{figure*}
    \centering
    \includegraphics[scale=0.3]{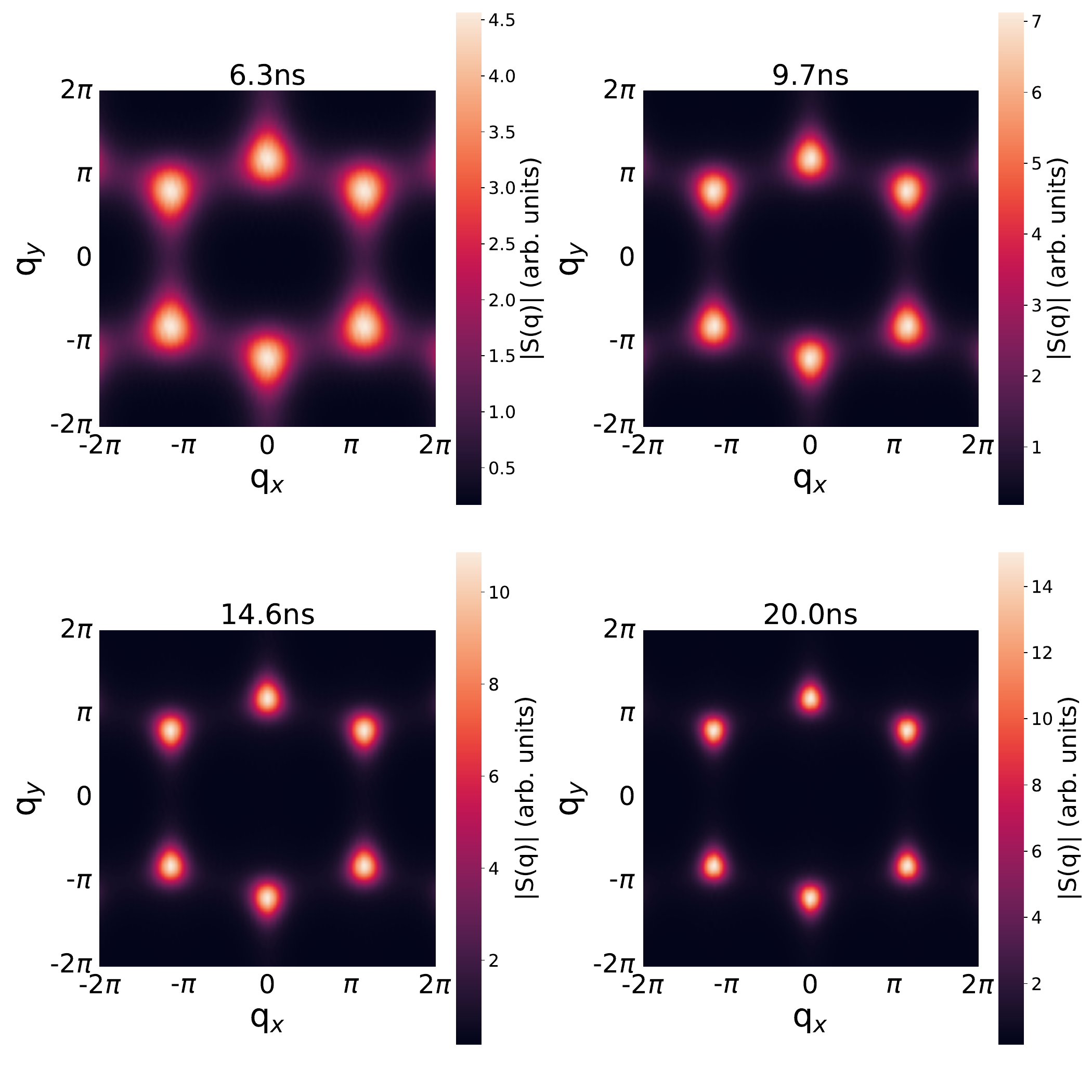}
    \caption{\textbf{SF of a $36\times 36$ triangular lattice with seperate scales.} The SF depicted in Fig.~\ref{fig5}a with an unnormalized scale showcasing the features in the Fourier space and the concentration of the peaks with increasing annealing time.}
\end{figure*}
\clearpage
\begin{figure*}
    \centering
    \includegraphics[scale=0.45]{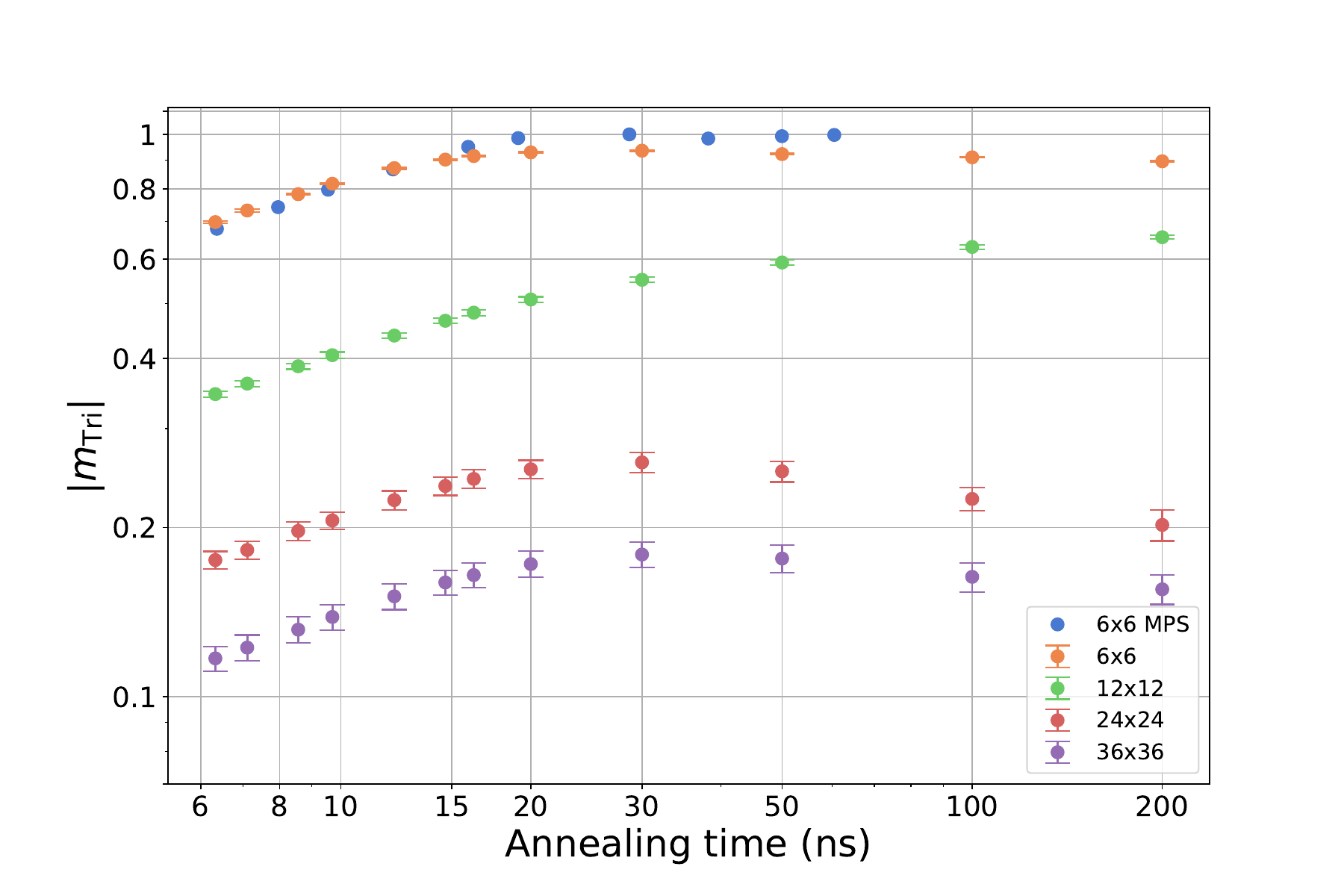}
    \caption{\textbf{Triangular OP scaling over a long annealing time.} Extending Fig.~\ref{fig4}a to \SI{200}{ns} where we can see the saturation of the OP. The MPS calculation is observed to saturate at a larger value indicating the rise of disorder in the QA at longer annealing times.}
\end{figure*}




\end{appendices}



\end{document}